\documentclass[aps,pra,twocolumn,groupedaddress,floatfix,showpacs]{revtex4}
\usepackage{amssymb,amsmath}
\usepackage{graphicx}
\usepackage{subfigure}
\usepackage[english]{babel}
\usepackage{float}
\usepackage{color}
\usepackage{booktabs}
\usepackage{blindtext}
\usepackage{comment}
\usepackage[symbol]{footmisc}

\usepackage{tikz,pgf}
\usepackage{soul,xcolor}
\setstcolor{red}
  
 \newcommand{\ket}[1]{\left|#1\right\rangle}

\begin{document}

\title{Manipulation of Multimode Squeezing in a Coupled Waveguide Array}

\author{S. Rojas-Rojas$^{1,2}$}
\author{E. Barriga$^3$}
\author{C. Mu\~noz$^3$}
\author{P. Solano$^4$}
\author{C. Hermann-Avigliano$^{2,5}$} \thanks{Corresponding author: carla.hermann@uchile.cl}
\affiliation{$^1$ Departamento de F\'isica, Universidad de Concepci\'on, 160-C Concepci\'on, Chile}
\affiliation{$^2$Millennium Institute for Research in Optics (MIRO), Concepción, Chile}
\affiliation{$^3$Departamento de F\'{\i}sica,  Facultad de Ciencias, Universidad de Chile,
Santiago, Chile}
\affiliation{$^4$Department of Physics, MIT-Harvard Center for Ultracold Atoms, and Research Laboratory of Electronics, Massachusetts Institute of Technology, Cambridge, Massachusetts 02139, USA}
\affiliation{$^5$Departamento de F\'{\i}sica,  Facultad
de Ciencias F\'isicas y Matem\'aticas, Universidad de Chile,
Santiago, Chile}

\pacs{03.67.Bg, 42.82.Et, 42.50.Dv, 42.50.Ex}

\begin{abstract}
We present a scheme for generating and manipulating three-mode squeezed states with genuine tripartite entanglement by injecting single-mode squeezed light into an array of coupled optical waveguides. We explore the possibility to selectively generate single-mode squeezing or multimode squeezing at the output of an elliptical waveguides array, determined solely by the input light polarization. We study the effect of losses in the waveguides array and show that quantum correlations and squeezing are preserved for realistic parameters. Our results show that arrays of optical waveguides are suitable platforms for generating multimode quantum light, which could lead to novel applications in quantum metrology.   
\end{abstract}

\maketitle

\section{Introduction}

Quantum metrology benefits from quantum resources to enhance measurements sensitivity beyond what is possible in classical systems~\cite{Giovannetti_2011}. Squeezed states are a remarkable example of quantum resources. These are states of light with reduced uncertainty in one of the quadratures of the electric field~\cite{SL-Knight}. Their potential applications in the fields of quantum metrology and quantum information have greatly increased in recent decades~\cite{PhysRevX,PhysRevLett,Nature-pht-circuits,Crespi_2012}, making possible the detection of gravitational waves with enhanced sensitivity~\cite{LIGO}.

Squeezed states can be generalized to multiple spatial modes of the electromagnetic field, for which noise reduction below the standard quantum limit (SQL) occurs for a linear combinations of the field?s quadratures. Multimode squeezed states have promising applications in multi-parameter quantum metrology~\cite{2018PhRvL.121m0503G,2017PhRvL.119m0504P,2016PhRvA..94e2108R}, multi-channel communication, and multi-channel quantum imaging~\cite{2009SciColorEntanglement,2004OptL...29..703S,2007PhRvL..98s0503S,2001Natur.414..413D}. Two-mode squeezed states were recently implemented surpassing the SQL for phase measurements~\cite{Anderson:17}. Three-mode squeezed states were obtained via spontaneous parametric six-wave mixing in an atomic-cavity system~\cite{2016Nat6WM}. However, the generation and manipulation of multimode-squeezed states remains difficult, partially because it requires highly non-linear processes. Their practical use presents challenges that would be easier to overcome by the manipulation of quantum light with linear optics.

Periodic optical structures, such as photonic crystals, are promising platforms to manipulate light~\cite{Kittel_1996,Joannopoulos_2008}. At low optical power, they behave as linear optical systems that allow for the modification of light propagation. These platforms have been extensively studied in the context of classical optics, but only recently they have enabled the manipulation of non-classical light~\cite{Rai,2014JOSAB..31..878R,2008PhRvL.101s3902L,2018SciA....4.3174T,2010Sci...329.1500P,2018SPIE10659E..0LK,2012NatPh...8..285A}.

Our research focuses on the theoretical study of squeezed light propagation, control, and manipulation in evanescently coupled waveguide arrays with linear response. We seek to understand how quantum features, such as entanglement and squeezing,  propagate in these structures. In particular, we study the propagation of quantum light in linear arrays of two and three waveguides. For the former case, we find a perfect analogy with the well-studied effect of a beam splitter \cite{Barnett}. In the case of a three-waveguides array we find that by injecting three single-mode squeezed states, the field evolves into a three-mode squeezed state with a rather simple experimental scheme. We also show that such evolution generates genuine multipartite entanglement. For specific coupling parameters of the waveguides it is possible to choose the output light to be in a multimode or single-mode squeezed state depending on the input light polarization. We consider the effect of losses in the photonic array in order to evaluate the feasibility of the proposed scheme, finding that squeezing and entanglement between waveguides are well preserved in typical experimental configurations. Our results suggest that photonic waveguides arrays are a reliable platform for creating entanglement and  three-mode squeezing, with potential scalability to higher-order multimode quantum states generation. This offers new strategies to control highly non-linear states of light with linear optics, which can potentially impact the scalability of quantum optics experiment with photonic systems.

The paper is organized as follow. We first study the scenario of two evanescently coupled waveguides in Sec. \ref{S_TMG}, and find that in such optical dimer the propagating state varies from two single-mode squeezed states to a two-mode squeezed state, known in the literature as \textit{two-mode squeezed Gaussons}~\cite{Barnett}. We then explore, in Sec. \ref{S_trimer}, the propagation of quantum light in a linear array of three waveguides, or optical trimer. Multipartite entanglement generation and propagation in optical dimers and trimers is studied in Sec. \ref{S_entanglement}. In Sec. \ref{S_polarization}, we show how light polarization serves to manipulate the order of multimodality of the output state. Finally, in Sec. \ref{S_losses}, we study the effect of losses in the system and conclude that quantum features are preserved in a realistic scenario. We present our conclusions and remarks in Sec. \ref{S_conclusions}.

\begin{figure}[!bt]
 \centering
 \includegraphics[width=.45\textwidth]{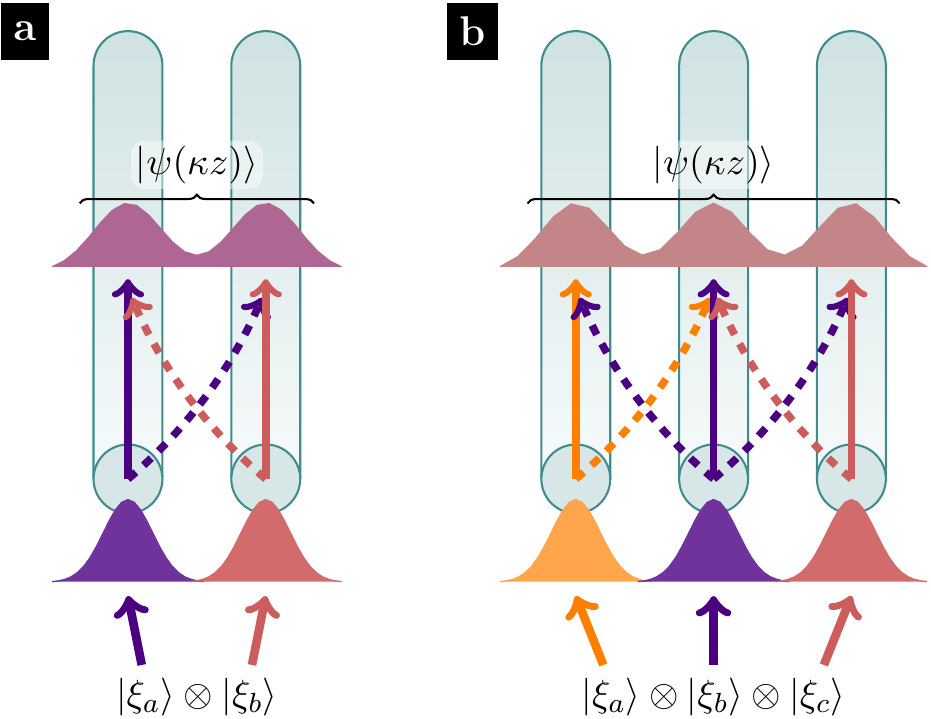}
 \caption{
 Squeezing distribution on an optical dimer (a) and trimer (b). Squeezed states defined by their squeezing parameter $\xi_i$ are coupled into the waveguides input. As the state propagates, light is evanescently coupled to neighboring waveguides, as depicted by the dashed arrows. The field evolution results in a multimode correlated state.}\label{fg:esquema}
\end{figure}

\section{Two-mode squeezing in an optical dimer}\label{S_TMG}

The propagation of light in a linear system with two input and two output ports [see Fig. \ref{fg:esquema} (a)] is described by two modes, $a$ and $b$, and an evolution operator $U$ that coherently couples both modes as
\begin{equation}\label{generalU}
U=\exp\left \{\gamma a^\dagger b-\gamma^*ab^\dagger\right \},
\end{equation}
with $\gamma=\theta\exp(i\delta)$, where $\theta$ and $\delta$ are the amplitude and phase of the complex coupling between modes. An initial input state $\ket{\psi_0}$ evolves as it propagates through the system, leading to the state $\ket{\psi}=U^{-1} \ket{\psi_0}$, using the notation from Ref. \cite{Barnett} for simplicity, and considering the fact that $U^{-1}$ differs from $U$ only in the sign of $\theta$. 

We study the particular case of an input field with two single-mode squeezed states
\begin{equation}
 \ket{\psi_0}=S_aS_b\ket{0,0},
\end{equation}
where the operators $S_a$ and $S_b$ are the single-mode squeezing operators for modes $a$ and $b$ defined by
\begin{equation}
S_j=\exp\left \{\frac{1}{2}(\xi^*_ja_j^2-\xi_j(a_j^\dagger)^2)\right \} ,
\label{S_operator}
\end{equation}
with $j=a,b$ and their respective squeezing parameter $\xi_{j}=r_j\exp(i\mu_j)$ a general complex number.

Operator $U$ can be inserted and removed to the left of the vacuum state, meaning that a rotation of the vacuum is again the vacuum, thus
\begin{equation}
 \ket{\psi}=U^{-1}S_aS_bU\ket{0,0}.
\end{equation}
As a result, we obtain a general squeezed state $\ket{\psi}$ for two modes
\begin{equation}
\label{2mgausson1}
 \ket{\psi}=\exp\left \{\frac{1}{2}(Z^*_aa^2+Z^*_bb^2+Z_{ab}b^\dagger a^\dagger- H.c.)\right \} \ket{0,0},
\end{equation}
where $Z_a$ and $Z_b$ are the coefficients of single-mode squeezing, and $Z_{ab}$ corresponds to the coefficient for two-mode squeezing. Depending on the value of these coefficients the state can vary from two single-mode squeezed states ($Z_{a}\vee Z_{b}\neq0$ and $Z_{ab}=0$) to a sole two-mode squeezed state ($Z_{a}=Z_{b}=0$ and $Z_{ab}\neq 0$). The term sole two-mode squeezing describes the situation where the noise of linear combination of the modes quadratures is below the vacuum noise and all single-mode quadratures noise is equal or higher than vacuum noise.For our particular case of $\ket{\psi_0}$, and using the unitary transformations
\begin{subequations}
 \begin{align}
\label{eq:dim_a}
 &U^{-1}aU=a\cos{\theta}+\exp{(i\delta)}b\sin{\theta},\\
 \label{eq:dim_b}
 &U^{-1}bU=b\cos{\theta}-\exp{(-i\delta)}a\sin{\theta},
 \end{align}
\end{subequations} 
the squeezing coefficients are
\begin{subequations}
 \begin{align}
  \label{eq:dimZ_a}
  Z_a& = \xi_a \cos^2(\theta)+\xi_b\exp(2i\delta) \sin^2(\theta)\,,\\
    \label{eq:dimZ_b}
  Z_b& = \xi_a \exp(-2i\delta)\sin^2(\theta)+\xi_b \cos^2(\theta)\,,\\
    \label{eq:dimZ_c}
  Z_{ab}&=-2\cos(\theta)\sin(\theta)[\xi_a\exp(-i\delta)-\xi\exp(i\delta)]\,.
 \end{align}
\end{subequations}

 From these equations we can identify two extreme cases. When $Z_{ab}=0$ and $Z_{a}\wedge Z_{b}\neq 0$ we get two single-mode squeezed states. On the other hand, when $Z_a=Z_b=0$ and $Z_{ab}\neq 0$ we get a sole two-mode squeezed state. The conditions for realizing both extreme states are summarized in Table \ref{tabla1}.

\begin{table}[htbp]
		\centering
\caption{Conditions to generate two single mode squeezed states or a sole two-mode squeezed state.}
		\label{tabla1}
		\begin{tabular}{cc}
			\hline
\multicolumn{2}{c}{Two single-mode squeezing} \\
			\hline
			Phases & $2\delta+\mu_a-\mu_b=2n\pi$  \\
			Squeezing strengths & $r_a=r_b$ \\
			$\theta$  & --- \\
			\hline
\multicolumn{2}{c}{Sole two-mode squeezing} \\
			\hline
			Phases & $2\delta+\mu_a-\mu_b=(2n+1)\pi$ \\
			Squeezing strengths & $r_a=r_b$ \\
			$\theta$   & $\theta=(2 m+1)\pi/4 $\\
			\hline
		\end{tabular}
		\label{table1}
	\end{table}
		
The unitary transformations (\ref{eq:dim_a}) and (\ref{eq:dim_b}) show that a quantum beam splitter is a particular case of the operator $U$ acting on two modes~\cite{Barnett}.

We are interested in a system of two evanescently coupled optical waveguides, i.e. an optical dimer [see Fig. \ref{fg:esquema} (a)]. The interaction between waveguides in this system is described by the Hamiltonian
\begin{equation}
H=-\hbar \kappa (a^\dagger b+ab^\dagger),
\end{equation}
where $\kappa$ is the coupling constant between the waveguides~\cite{Santiago,PhysRev.162.1256,2010Sci...329.1500P,PhysRev.175.286}. We notice that the evolution operator of such a Hamiltonian corresponds to the operator $U$ in the particular case of $\delta=\pi/2$,
\begin{equation}
U=\exp\left \{i\theta(a^\dagger b+ab^\dagger)\right \},
\end{equation}
with $\theta=\kappa z$, where $z$ is the propagation distance. The light state evolves along the direction of propagation $z$, hence, $\kappa z$ is the dimensionless parameter controlling the evolution of the state.  The squeezing coefficients of the field propagating through the optical dimer are a particular case of Eqs. (\ref{eq:dimZ_a})-(\ref{eq:dimZ_c}), now in therms of $\kappa z$
\begin{subequations}\label{eq:dim}
 \begin{align}
  \label{coefDimer}
  Z_a(\kappa z)& = \xi_a \cos^2(\kappa z)-\xi_b \sin^2(\kappa z)\,,\\
  Z_b(\kappa z)& = -\xi_a \sin^2(\kappa z)+\xi_b \cos^2(\kappa z)\,,\\
  Z_{ab}(\kappa z)&=2i(\xi_a+\xi_b)\cos(\kappa z)\sin(\kappa z)\,.
 \end{align}
\end{subequations}

Notice that for particular propagation distances we can obtain either two single-mode squeezed states [$\kappa z=n\pi/2$], or a sole two-mode squeezed state [$\kappa z=(2n+1)\pi/4$ and $\xi_a=\xi_b$.\\

\begin{figure}[!bt]
\includegraphics[width=.47\textwidth]{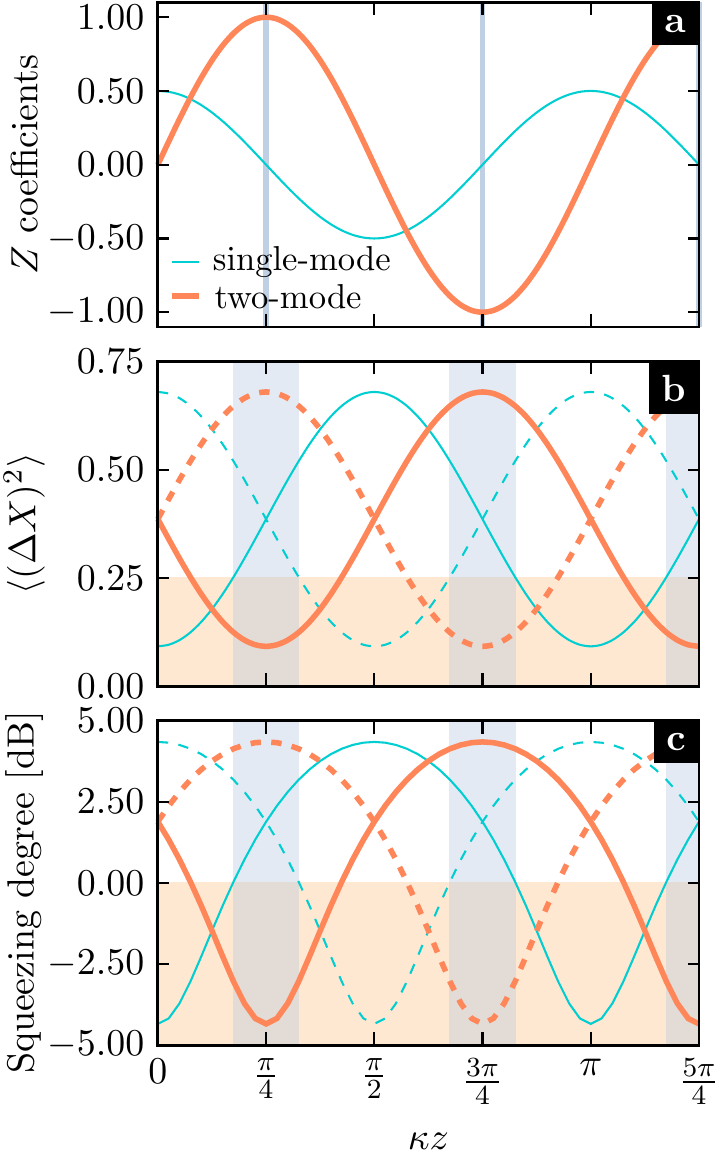}
 \caption{Evolution of the field along the optical dimer when two single-mode squeezed states with $\xi_a=\xi_b=0.5$ are injected separately into the waveguides. 
 (a) $Z_a=Z_b$ coefficients (thin cyan curve) and $|Z_{ab}|$ (thick orange curve) as a function of $\kappa z$. The vertical gray lines show the points where  the  all  the  single-mode  coefficients  ($Z$)  vanish (b) Variances of the quadratures as a functions of $\kappa z$ for both single-mode quadratures variances $\langle(\Delta X^{(a,b)}_1)^2\rangle $ and $\langle(\Delta X^{(a,b)}_2)^2\rangle $ (thin solid and dashed cyan curve), as well as both two-mode quadrature variances $\langle(\Delta X^{2M}_1)^2\rangle $ and $\langle(\Delta X^{2M}_2)^2\rangle $ (thick solid and dashed orange curves). (c) Squeezing degree as a function of $\kappa z$ for the same quadratures as in (b). Squeezing is observed when the curves in (b) and (c) are within the orange region. The vertical gray bands in (b) and (c) show the regions where sole two-mode squeezing is observed.} 
 \label{fg:dima}
\end{figure}

\begin{figure*}[!tb]
  \includegraphics[width=.8\textwidth]{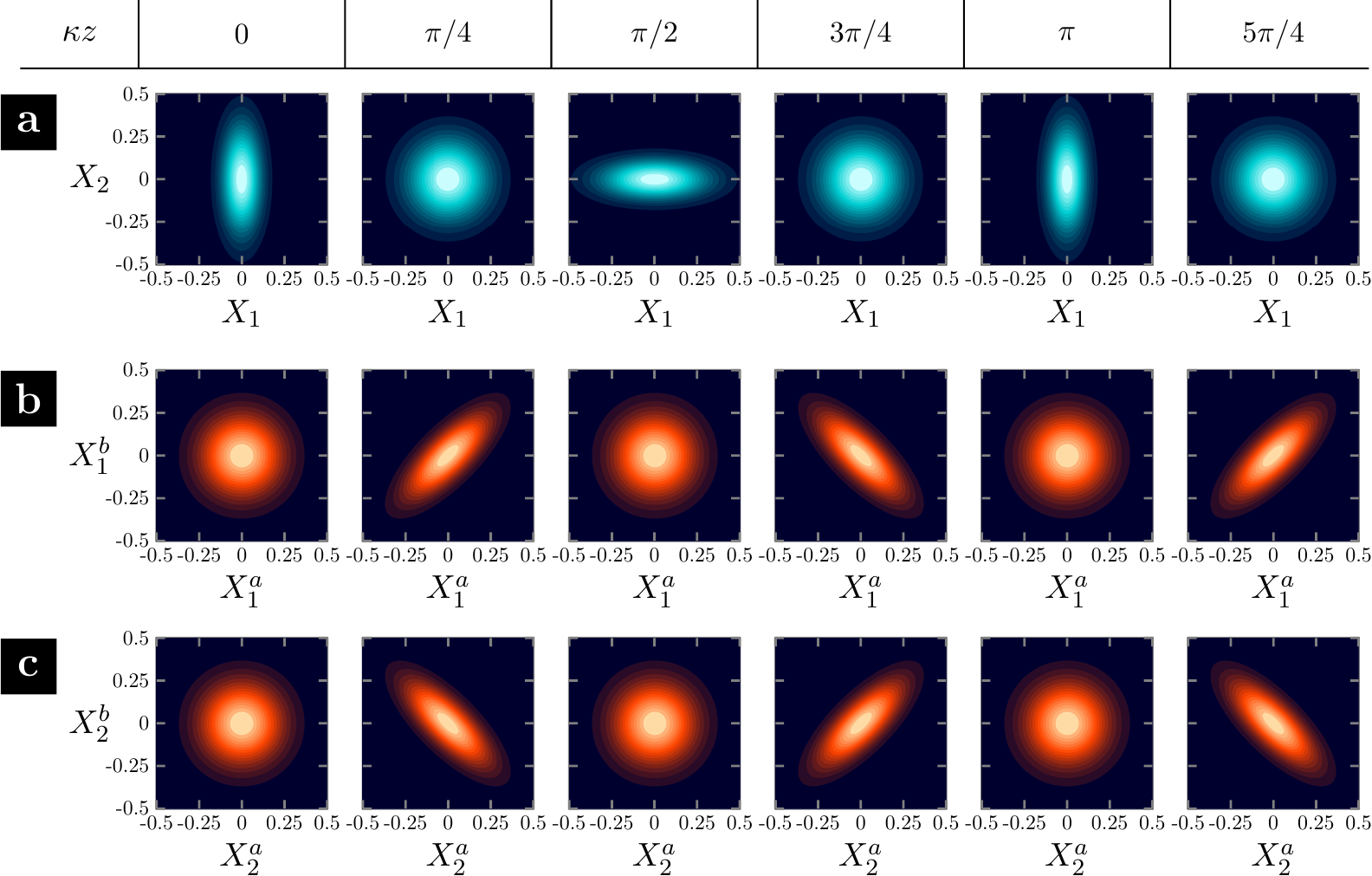}
 \caption{Evolution of the Wigner function of the reduced single-mode and two-mode states along an optical dimer. (a) Wigner function of the reduced single-mode states on waveguides $a$ and $b$. (b) and (c) are the marginal distributions $|\psi(X_1^{a},X_1^{b})|^2$ and  $|\psi(X_2^{a},X_2^{b})|^2$ of the two-mode Wigner function. The initial state and the color code are the same as Fig. \ref{fg:dima}.}\label{fg:wigdim}
\end{figure*}

To study the squeezing evolution in an optical dimer we define the single-mode quadratures for mode $a$ as
\begin{subequations}
\begin{align}
 X^{a}_1&=\frac{1}{2}(e^{-i\phi_0}a+e^{i\phi_0}a^\dagger)\,,\\
 X^{a}_2&=\frac{1}{2i}(e^{-i\phi_0}a-e^{i\phi_0}a^\dagger)\,
\end{align}
 \end{subequations}
and likewise for mode $b$, where $\phi_0$ is the angle that defines the measured quadratures (typically the squeezed and anti-squeezed ones).

Two-mode squeezed states exhibit squeezing in a superposition of quadratures from both modes. The generalized two-mode quadratures are~\cite{Alberto}
\begin{subequations}\label{eq:x2m}
\begin{align}
 X_1^\text{2M}&=\frac{1}{2^{3/2}}\left[e^{-i\phi}(a+b)+e^{i\phi}(a^\dagger+b^\dagger)\right],\\
 X_2^\text{2M}&=\frac{1}{2^{3/2}i}\left[e^{-i\phi}(a+b)-e^{i\phi}(a^\dagger+b^\dagger)\right],
\end{align}
\end{subequations}
where $\phi$ corresponds to the angle where we expect to observe squeezing, in analogy to the single-mode case. 

The degree of squeezing is obtained from the variance of the quadratures as
\begin{equation}
S(dB)=10\log_{10} \left[ \frac{\langle(\Delta X)_{sq}^2\rangle}{\langle(\Delta X)_{ch}^2\rangle} \right],
\end{equation}
where $\langle(\Delta X)_{sq}^2\rangle $ is the variance of the field in the (anti-) squeezed quadrature and $\langle(\Delta X)_{ch}^2\rangle=1/4 $ is the variance of a coherent state. The variance of different quadratures can be easily calculated and they are given in Appendix \ref{ap:quadrature}.

Figure \ref{fg:dima} shows the evolution of the input field propagating through an optical dimer. We assume a particular input state $|\psi_0\rangle$, where the squeezing parameters $\xi_a=\xi_b$ are real. In this case the coefficient $Z_{ab}(\kappa z)$ from Eq. (\ref{eq:dimZ_c}) is always imaginary, meaning that a proper quadrature to measure two-mode squeezing would be $\phi=(2n+1)\pi/4$. In Fig. \ref{fg:dima} (a) we plot the evolution of the single-mode squeezing coefficients $Z_{a}=Z_{b}$ as well as the two-mode squeezing coefficient $Z_{ab}$. For $\kappa z=0$, the input state has the same single-mode squeezing in both waveguides and $Z_{ab}=0$. For $\kappa z=(2n+1)\pi/4$, we get $Z_a=Z_b=0$ and $Z_{ab}$ is maximum, meaning a sole two-mode squeezed state. Figure \ref{fg:dima}(b) shows the evolution of the single mode and generalized two-mode quadrature variances. We notice that for the single-mode quadrature variances the squeezing is lost before $Z_a=Z_b=0$, because $Z_{ab}\neq0$ (see Appendix \ref{ap:quadrature} for the analytic expressions). This feature is represented by the gray bands in Fig. \ref{fg:dima}.  

The degrees of squeezing of each single-mode squeezed input state are transferred to a sole two-mode squeezed state, and vice versa, as Fig. \ref{fg:dima}(c) shows. Notice that we can analogously inject a two-mode squeezed state into the optical dimer and get two independent single-mode squeezed states at the output, as shown in detail in Appendix \ref{ap:TMSS}.

If we inject a single-mode squeezed state only into waveguide mode $a$, leaving $b$ in the vacuum state, then after a propagation distance of $\kappa z=(2n+1)\pi/2$ all the squeezing is transferred from mode $a$ to $b$~\cite{Rai}. However, when squeezing is just injected in one of the waveguides, it is impossible to generate a sole two-mode squeezed state.

The transfer of squeezing between modes can be visualized using the Wigner function~\cite{HarocheBook,Braunstein}, a quasi-probability distribution of the fields in the quadrature space. The Wigner function of a vacuum state gives a symmetric Gaussian distribution, while a squeezed state shows a narrowing of the distribution along a given direction. We compute the Wigner functions for the reduced state of each waveguide, as well as for the two-mode (bipartite) state. The Wigner function of two-mode states $W^{2M}$ is defined by four variables, namely a pair of quadratures on each mode. In order to make its visualization possible we compute the marginal distributions of the complete Wigner function, which reflects the correlations of fields propagating through optical array. This gives a quasi-probability distribution as a function of a subset of two out of four modes. We consider the marginal distributions for two pairs of cross-correlated quadratures
\begin{align}
 |\psi(X_1^{a},X_1^{b})|^2&=\int d X_2^{a} dX_2^{b}\, W^{2M} (X_1^{a},X_2^{a},X_1^{b},X_2^{b})\\
  |\psi(X_2^{a},X_2^{b})|^2&=\int d X_1^{a} dX_1^{b}\, W^{2M} (X_1^{a},X_2^{a},X_1^{b},X_2^{b}).
\end{align}

Figure \ref{fg:wigdim} shows the Wigner functions of the single- and the two-mode states, considering the same parameters as in Fig. \ref{fg:dima}. At the input ($\kappa z=0$) the single-mode states of each waveguide exhibit a Wigner function squeezed on the $X_1$ quadratures ($\xi_a=\xi_b$). As the state propagates ($\kappa z=\pi/4$), the single-mode squeezing is lost. In contrast, the Wigner function for the two-mode state is symmetric at the input and, as the state propagates, $|\psi(X_1^{a},X_1^{b})|^2$ and $|\psi(X_2^{a},X_2^{b})|^2$ become squeezed, exhibiting correlation and anticorrelation between the cross-correlated quadratures.

\section{Three-mode squeezing in an optical trimer}
\label{S_trimer}

A linear optical system with three input and three output ports, as Fig. \ref{fg:esquema} (b) shows, is described by three modes, $a$, $b$, and $c$, and the evolution operator
\begin{equation}
\label{3rotation}
U_{\text{T}}=\exp\left \{(\alpha^*ab^\dagger-\alpha ba^\dagger+\beta^*bc^\dagger-\beta cb^\dagger)\right \},
\end{equation}
that coherently couples all three modes. $\alpha=\theta_{\alpha}e^{i\delta_{\alpha}}$ and $\beta=\theta_{\beta}e^{i\delta_{\beta}}$ are complex numbers that characterize the coupling strength between neighboring waveguides.

We study the particular case of injecting a single-mode squeezed states into each input port, $|\phi_0\rangle_{\text{T}}=S_aS_bS_c\ket{0,0,0}$, with the single-mode squeezing operator defined in Eq. (\ref{S_operator}), in analogy with Sec.\ref{S_TMG}. Light propagates through the waveguide array evolving into the state 
\begin{equation}
\label{3mgausson2}
 \ket{\Psi}_{\text{T}}=U_{\text{T}}^{-1}S_aS_bS_cU_{\text{T}}\ket{0,0,0}.
\end{equation}
After propagation, we obtain a general state with mixed characteristics of single-mode, two-mode, and three-mode squeezing, namely~\cite{def.3MSS},
\begin{equation}
\label{3mgausson}
\begin{split}
 \ket{\Psi}_{\text{T}}=&\exp\left \{\frac{1}{2}(T^*_aa^2+T^*_bb^2+T^*_cc^2\right.\\
 &\left.+T_{ab}a^\dagger b^\dagger+T_{ac}a^\dagger c^\dagger+T_{bc}b^\dagger c^\dagger- H.c.)\right \} \ket{0,0,0}
 \end{split}
\end{equation}
The squeezing parameters $T_i$ and $T_{ij}$ (with $i,j=a,b,c$) have an analogous interpretation as the squeezing $Z$-coefficient in the optical dimer. In the case of $T_a\vee T_b\vee T_c\neq0$ and $T_{ab}=T_{bc}=T_{ac}=0$ we have sole single-mode squeezing. If $T_a=T_b=T_c=0$ and only one $T_{ij}\neq 0$ with the other pairwise coefficients equal to zero, we have sole two-mode squeezing. Finally, if $T_a=T_b=T_c=0$ and $T_{ab}\wedge T_{bc}\wedge T_{ac}\neq0$, we have sole three-mode squeezing. This comes from the definition of multimode-squeezing, where the uncertainty of the pairwise sum of quadratures is reduced for all the pairs of modes~\cite{def.3MSS}. As in Sec. \ref{S_TMG}, the term sole three-mode squeezing describes a field where all single-mode quadratures noise is equal or higher than vacuum noise.

The squeezing $T$ coefficients are directly calculated using the rotations
\begin{subequations}
\label{TranfUnitarias2}
\begin{align}
  U^{-1}  a   U &= \cos^2{\left( \frac{\theta}{\sqrt{2}}\right)}   a +\frac{e^{i\delta}}{\sqrt{2}}\sin{\left(\sqrt{2}\theta\right)}   b\nonumber\\
&+e^{2i\delta}\sin^2{\left(\frac{\theta}{\sqrt{2}}\right)}  c,\\
      U^{-1}  b   U & = -\frac{1}{\sqrt{2}}e^{-i\delta}\sin{\left(\sqrt{2}\theta\right)}  a + \cos{\left(\sqrt{2}\theta \right)}  b\nonumber\\
    &+\frac{1}{\sqrt{2}}e^{i\delta}\sin{\left(\sqrt{2}\theta\right)}  c,\\
      U^{-1}  c   U &=e^{-2i\delta} \sin^2{\left( \frac{\theta}{\sqrt{2}}\right)}  a -\frac{1}{\sqrt{2}}e^{-i\delta}\sin{\left(\sqrt{2}\theta\right)}  b\nonumber\\
    &+\cos^2{\left( \frac{\theta}{\sqrt{2}}\right)}   c,
\end{align}
\end{subequations}
where we assume equal coupling among waveguides, i.e., $\alpha=\beta$, for simplicity. The most general rotations are detailed in Appendix \ref{apendice}.

We study an optical trimer consisting of a linear array of three identical coupled waveguides, as Fig. \ref{fg:esquema}(b) shows. The interaction Hamiltonian of the system is~\cite{Rai,Santiago,PhysRev.162.1256,2010Sci...329.1500P,PhysRev.175.286} 
\begin{equation}
H=-\hbar \kappa (a^\dagger b+ab^\dagger+b^\dagger c +c^\dagger b).
\end{equation}
The evolution of the light propagating through the waveguides along the $z$ direction is governed by the coupling constant $\kappa$. An optical trimer is a particular case of $U_{\text{T}}$ on Eq.(\ref{3rotation}) for $\delta=\pi/2$. Then, the squeezing $T$-coefficients are
\begin{equation}\label{coeftrimero}
\begin{split}
T_a&= \xi_a\cos^4{\left( \frac{\kappa z}{\sqrt{2}}\right)}-\xi_b\frac{1}{2}\sin^2{\left(\sqrt{2}\kappa z\right)}+\xi_c\sin^4{\left(\frac{\kappa z}{\sqrt{2}}\right)} \\
T_b&= -\xi_a\frac{1}{2}\sin^2{\left( \sqrt{2}\kappa z\right)}+\xi_b\cos^2{\left( \sqrt{2}\kappa z\right)}-\xi_c\frac{1}{2}\sin^2{\left(\sqrt{2}\kappa z\right)} \\
T_c&= \xi_a \sin^4{\left( \frac{\kappa z}{\sqrt{2}}\right)}-\xi_b\frac{1}{2}\sin^2{\left( \sqrt{2}\kappa z\right)}+\xi_c\cos^4{\left( \frac{\kappa z}{\sqrt{2}}\right)} \\
T_{ab} &=i\xi_a \sqrt{2}\sin{\left( \sqrt{2}\kappa z\right)}\cos^2{\left( \frac{\kappa z}{\sqrt{2}}\right)}+i\xi_b\frac{1}{\sqrt{2}}\sin{\left(2\sqrt{2}\kappa z \right)}\\
&-i\xi_c\sqrt{2}\sin{\left( \sqrt{2}\kappa z \right)}\sin^2{\left( \frac{\kappa z}{\sqrt{2}}\right)} \\
T_{ac} &=\xi_a \frac{1}{2}\sin^2{\left( \sqrt{2}\kappa z\right) }+\xi_b\sin^2{\left( \sqrt{2}\kappa z\right)}+\xi_c\frac{1}{2}\sin^2{\left( \sqrt{2}\kappa z\right)} \\
T_{bc} &= -i\xi_a \sqrt{2}\sin{\left( \sqrt{2}\kappa z\right)}\sin^2{\left( \frac{\kappa z}{\sqrt{2}}\right)}+i\xi_b \frac{1}{\sqrt{2}}\sin{\left( 2\sqrt{2}\kappa z\right)}\\
&+i\xi_c\sqrt{2}\sin{\left( \sqrt{2}\kappa z\right)}\cos^2{\left( \frac{\kappa z}{\sqrt{2}}\right)}.
\end{split}
\end{equation}

\begin{figure}[!bt]
\includegraphics[width=.47\textwidth]{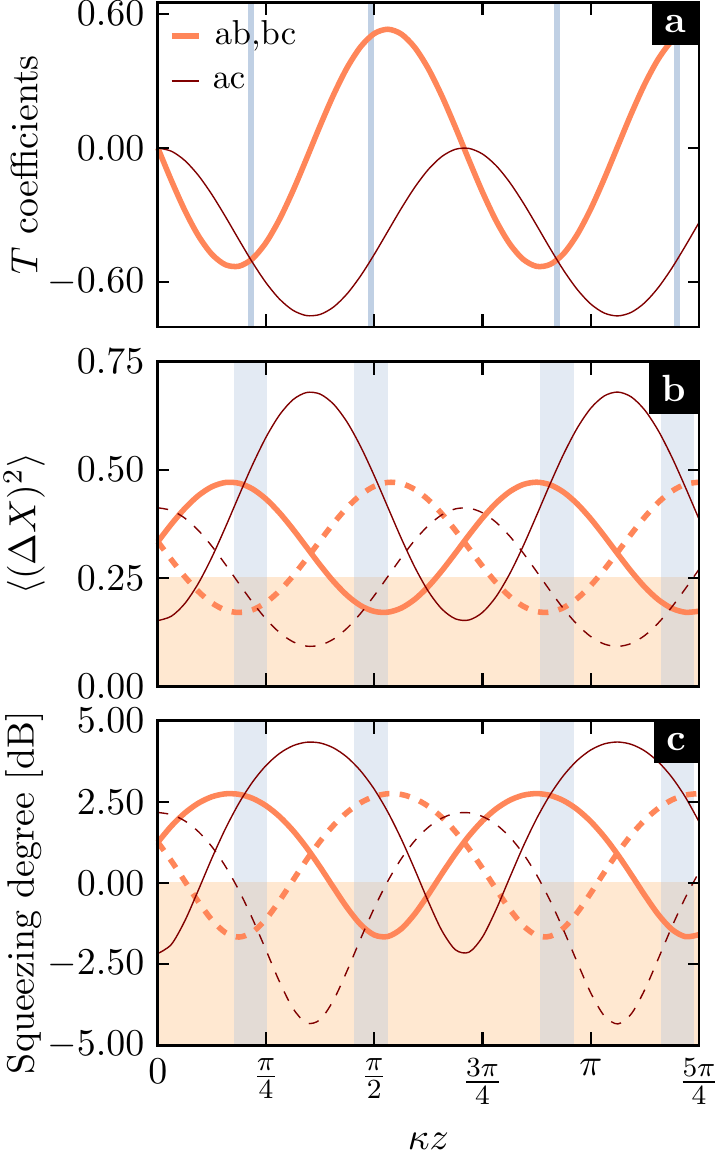}
 \caption{Evolution of the field along the optical trimer when three single-mode squeezed states, with $\xi_a=\xi_c=0.25$ and $\xi_b=0.5$, are injected into the waveguides. (a) Amplitude of the $T_{ab}=T_{bc}$ coefficients (thick orange curve) and the $T_{ac}$ coefficient (thin brown curve) as a function of the normalized propagation distance $\kappa z$. The vertical gray lines show the points where the all the single-mode coefficients ($T_j$) vanish. (b) Variances of the quadratures as a function of $\kappa z$ for the two-mode quadrature variances $X_1^{ab}=X_1^{bc}$ and $X_2^{ab}=X_2^{bc}$ (thick solid and dashed orange curves), as well as $X_1^{ac}$ and $X_2^{ac}$ (thin solid and dashed brown curves). (c) Squeezing degree as a function of $\kappa z$ for the same quadratures than in (b). Squeezing is observed when the curves in (b) and (c) are within the orange region. The vertical gray bands in (b) and (c) show the regions where sole three-mode squeezing is observed.} 
 \label{fg:trimer}
\end{figure}

\begin{figure}[!bt]
  \includegraphics[width=.48\textwidth]{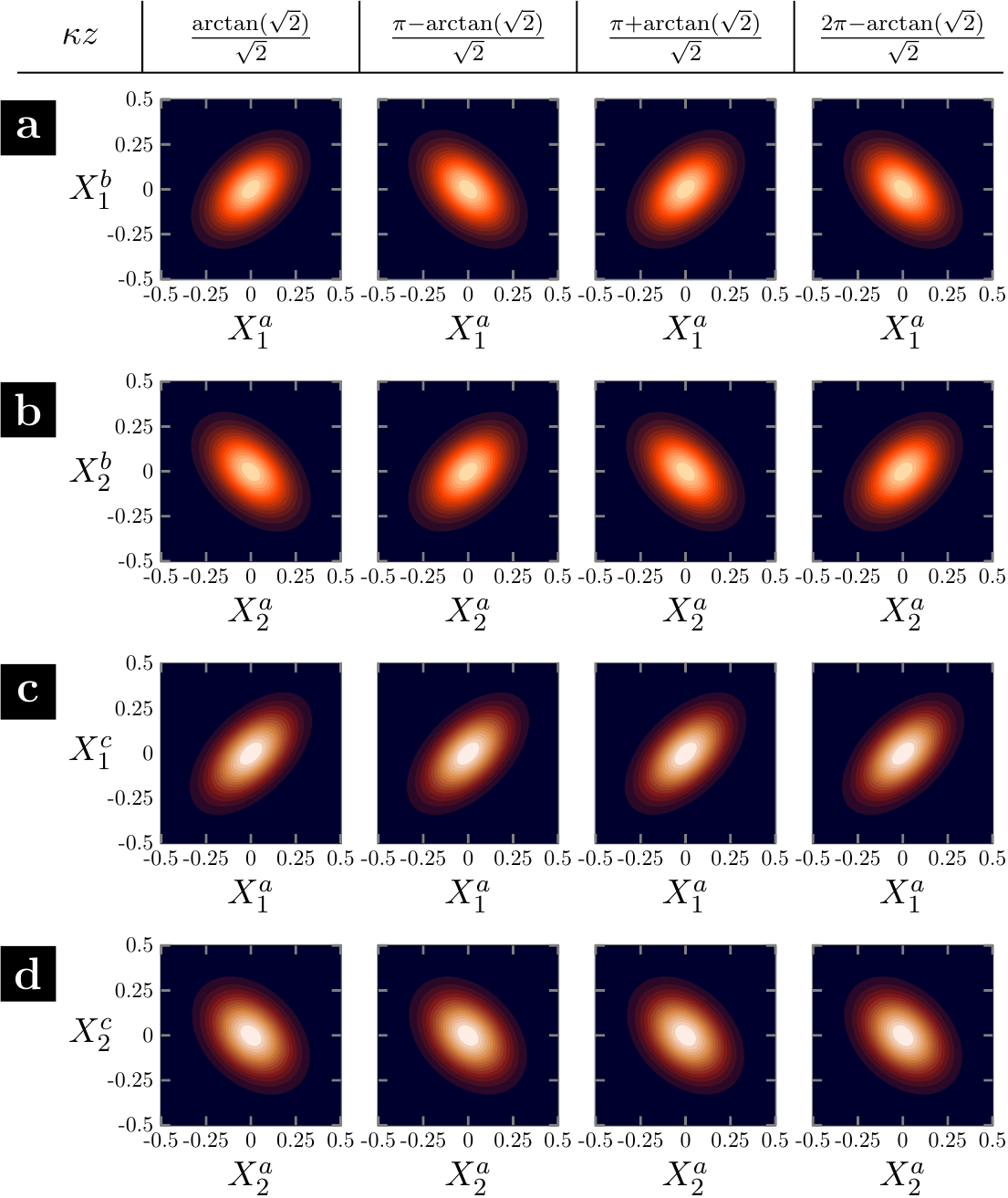}
 \caption{Evolution of the marginal distributions from the Wigner function of different pairs of modes. (a) Marginal distribution $|\psi(X_1^{a},X_1^{b})|^2=|\psi(X_1^{b},X_1^{c})|^2$. (b) Marginal distribution $|\psi(X_2^{a},X_2^{b})|^2=|\psi(X_2^{b},X_2^{c})|^2$. (c) Marginal distribution $|\psi(X_1^{a},X_1^{c})|^2$. (d) Marginal distribution $|\psi(X_2^{a},X_2^{c})|^2$. The initial state and the color code are the same as in Fig.\ref{fg:trimer}.}\label{fg:wigtrim}
\end{figure}

These equations describe a general solution for the propagation of squeezed light through an optical trimer with equal coupling coefficients. Three-mode squeezed states are generated for an input state $|\phi_0\rangle_{\text{T}}$ with real squeezing parameters $\xi_a=\xi_c=\xi_b/2$. Figure \ref{fg:trimer} shows the evolution of the input field propagating through an optical trimer. Figure. \ref{fg:trimer}(a) shows $T_{ab}=T_{bc}$ and $T_{ac}$ as a function of the propagation parameter $\kappa z$. From Eq. (\ref{coeftrimero}) we can easily verify that for $\kappa z=n\pi/\sqrt{2}$, $T_{ab}=T_{ac}=T_{bc}=0$, as Fig. \ref{fg:trimer}(a) shows, while $T_a$, $T_b$, and $T_c$ are different from zero [not shown in Fig. \ref{fg:trimer}(a)]. For $\kappa z=(2n+1)\pi/{2\sqrt{2}}$, the coefficients $T_{ab}$ and $T_{bc}$ vanish, while $T_{ac}$ does not. However, this is not a sole two-mode squeezed state, since $T_a$, $T_b$, and $T_c$ are also different from zero. In order to get a sole three-mode squeezed state we need $T_a=T_b=T_c=0$. This happens for angles $\kappa z_1=\left[n\pi+\arctan(\sqrt{2})\right]/\sqrt{2}$ and $\kappa z_2=\left[n\pi-\arctan(\sqrt{2})\right]/\sqrt{2}$, for the particular case of input states with $\xi_a=\xi_c=\xi_b/2$ [vertical blue lines in Fig. \ref{fg:trimer} (a)]. 

Figures \ref{fg:trimer} (b) and (c) show the evolution of the variances of the generalized two-mode quadratures and the degree of squeezing for each possible pair of modes, $ab$, $bc$, and $ac$. Both the variances and the squeezing degree for the combined modes $ab$ and $bc$ are always equal. Sole three-mode squeezing is observed within the gray bands, where single-mode squeezing is absent from all the waveguides, in analogy to Fig. \ref{fg:dima} in Sec. \ref{S_TMG}. We notice that a reduction or amplification of noise (relative to vacuum noise) can appear for a particular sum of quadratures when considering two modes with single-mode squeezing [see the brown curves in Figs. \ref{fg:dima} (b) and (c) at $\kappa z=0$]. This is the result of adding two fields with reduced or increased uncorrelated noise and does not signify a correlation between them. However, when the single-mode squeezing is zero in all modes ($T_{a}=T_{b}=T_{c}=0$) we can guarantee multi-mode squeezing with genuine quantum correlation.

We can visualize the evolution of the three-mode squeezed state using the Wigner function following the analysis in Sec. \ref{S_TMG}. Figure \ref{fg:wigtrim} shows the 
the marginal distributions of the Wigner function for different combinations of two-mode states. In particular, it shows the Wigner function at the propagation distances $\kappa z$ where squeezing is observed in all the combinations of two-modes but not in each individual mode, meaning a sole three-mode squeezed state.

If squeezing is injected only into waveguide $a$ we obtain complete transfer of squeezing from waveguide $a$ to $c$ for $kz=\pi/\sqrt(2)$~\cite{Rai}. When squeezing is injected just in the middle waveguide, squeezing is transferred to waveguide $a$ and waveguide $c$, for $kz=\sqrt(2)\pi/4$. This is a general two-mode squeezed state with $T_a=T_c=-\xi_b/2$ and $T_{ac}=-\xi_b$, such as the state in Eq. (\ref{2mgausson1}), while all the other $T$ coefficients vanish. 

\section{Entanglement generation and propagation}
\label{S_entanglement}

\begin{figure}[!bt]
 \includegraphics[width=0.5\textwidth]{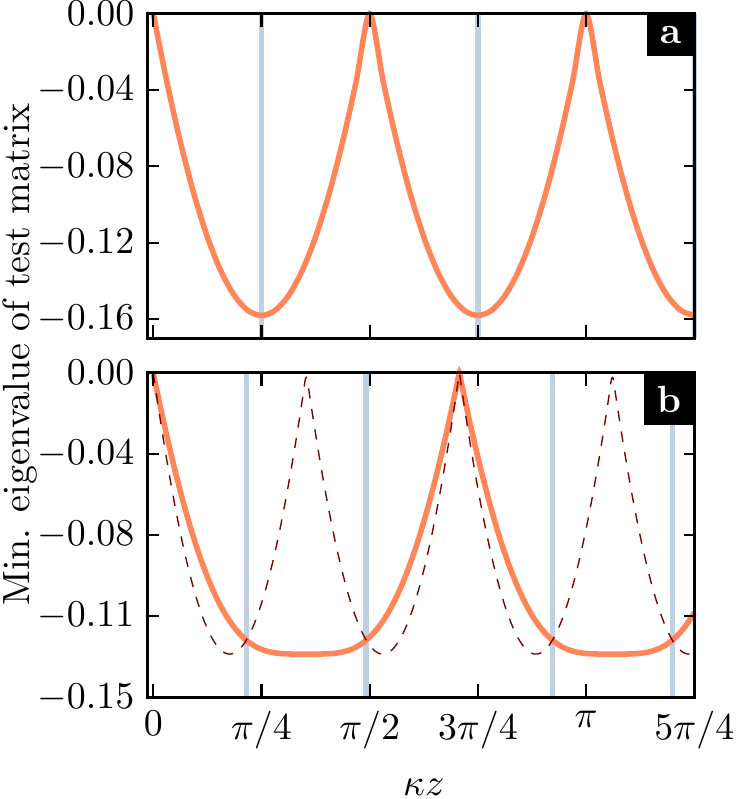}
  \caption{Entanglement properties of the light propagating through the waveguides arrays ($\kappa z$). Negative eigenvalues correspond to entangled states. (a) Minimum eigenvalue of the matrix $\bar V^{(2)}_j-\frac{i}{4}\,\Lambda$, for both $j=a,b$ (c) Minimum eigenvalue of matrices $\bar V^{(3)}_j-\frac{i}{4}\,\Lambda$ for $j=a$ and  $j=c$ (orange thick curve), and $j=b$ (brown thin dashed curve). The vertical gray lines show the angles where sole two-mode} (a) and three-mode (b) squeezing is observed.\label{fg:ent}
\end{figure}

Multimode squeezed states show evident cross-correlations between quadratures of different modes. However, we cannot assume that these correlations are quantum in nature, i.e. entanglement. Even though squeezing and entanglement are closely related, to the point that two-mode squeezed states can be a physical realization of the ideal two-particle Einstein-Podolsky-Rosen (EPR) state \cite{Braunstein}, they do not have a direct correspondence. The defining property of an entangled state is its non-separability, meaning its density matrix cannot be represented as the external product of two density matrices. A particular test for non-separability, known as the Peres-Horodecki criterion~\cite{Peres1996,Horodecki1997}, tells that a state is non-separable if the partial transpose of its density matrix has negative eigenvalues. 

The Peres-Horodecki criterion can be extended to multipartite continuous-variable states as follows~\cite{cvnpt,Braunstein}. The Wigner function for Gaussian states of $N$ modes is characterized by a correlation matrix $V^{(N)}$, a $2N\times 2N$ matrix with elements defined by
\begin{equation}
V_{ij}^{(N)}={\rm tr}\left\lbrace\rho\,(\Delta\zeta_i\Delta\zeta_j+
\Delta\zeta_j\Delta\zeta_i)/2\right\rbrace,
\end{equation}
where $\zeta=(X^a_1~ X^a_2~ X^b_1~ X^b_2~...~ X^N_1~ X^N_2)$ is a 2$N$-dimensional vector operator that contains all the quadrature operators. In the particular case of zero mean value of the quadratures (such as vacuum state) $V_{ij}^{(N)}
=\langle(\hat\zeta_i\hat\zeta_j+\hat\zeta_j\hat\zeta_i)/2\rangle\,$. Under this approach the partial transposition of a continuous-variable state is simply a sign change of the momentum quadrature ($X_2$) of a subsystem. The partial transpose operation acts on the correlation matrix as $\bar V^{(N)}_j= \Gamma_j V^{(N)}\Gamma_j$, where $\Gamma_j$ are the transposition operator performing the sign change of the $X_2$ variable in subsystem $j$. The negative partial transpose criterion for continuous-variables says that a separable state satisfies the $N$-mode uncertainty relation even after partial transposition in site $j$, i. e.,
\begin{equation}\label{eq:gunc}
\bar V^{(N)}_j\  \geq\frac{i}{4}\,\Lambda\,,
\end{equation}
with $\Lambda$ the $2N\times 2N$ block matrix containing the values of commutators between all the possible pairs of quadratures from every subsystem ($\frac{i}{2}\Lambda_{ij}=\left[\zeta_i,\zeta_j\right]$). For example, in a bipartite system: 
\begin{equation*}
 \Lambda=
\left( \begin{array}{cc} J & 0 \\
0 & J
\end{array} \right)\;,\quad\quad
J=
\left( \begin{array}{cc} 0 & 1 \\
-1 & 0
\end{array} \right)\,.
\end{equation*}

The negative partial transpose criterion for continuous-variables is summarized from Eq. \eqref{eq:gunc} as follows: an $N$-modes state is entangled if $\left(\bar V^{(N)}_j-\frac{i}{4}\,\Lambda\right)$ has a negative eigenvalue. Figure \ref{fg:ent} shows the the minimum of the eigenvalues of the operators $\left(\bar V^{(N)}_j-\frac{i}{4}\,\Lambda\right)$ for the optical dimer (a) and trimer (b), as a function of the propagation distance $\kappa z$. The input states ($\kappa z=0$) have a minimum eigenvalue of zero, since our initial condition is a product of uncoupled squeezed states. As light propagates through the waveguides array, the minimum eigenvalues become negative, evidencing multimode entanglement, in agreement with the behavior of the squeezing parameters and the Wigner functions (see Figs. \ref{fg:dima}-\ref{fg:wigtrim}). 
  
We are particularly interested in the case of an optical trimer, where quantum correlations between three-mode continuous variables can be observed. The study of tripartite entanglement for arbitrary dimensions is still a subject of research~\cite{multicv}, due to the difficulties to determine how the entanglement is distributed among the parties. To determine the presence of genuine tripartite entanglement we show the full inseparability of the state. In general, these two concepts are not equivalent; however, for pure states, like the ones studied here, full inseparability implies genuine tripartite entanglement (see Ref.~\cite{Armstrong} and Supplementary Material in Ref.~\cite{Shalm}). In order to verify full inseparability, it is necessary to rule out all the possible partially separable forms, corresponding to all the combinations of bipartite subsystems. From  the negative partial transpose criterion for continuous-variables we can distinguish the following four scenarios \cite{casos}:
\begin{align*}
&{\rm I}.&\bar V^{(3)}_a\ngeq \frac{i}{4}\,\Lambda\,,
\bar V^{(3)}_b\ngeq \frac{i}{4}\,\Lambda\,,
\bar V^{(3)}_c\ngeq \frac{i}{4}\,\Lambda\,,\\
&{\rm II}.&\bar V^{(3)}_k\geq \frac{i}{4}\,\Lambda\,,
\bar V^{(3)}_m\ngeq \frac{i}{4}\,\Lambda\,,
\bar V^{(3)}_n\ngeq \frac{i}{4}\,\Lambda\,,
\nonumber\\
&{\rm III}.&\bar V^{(3)}_k\geq \frac{i}{4}\,\Lambda\,,
\bar V^{(3)}_m\geq \frac{i}{4}\,\Lambda\,,
\bar V^{(3)}_n\ngeq \frac{i}{4}\,\Lambda\,,
\nonumber\\
&{\rm IV}.&\bar V^{(3)}_a\geq \frac{i}{4}\,\Lambda\,,
\bar V^{(3)}_b\geq \frac{i}{4}\,\Lambda\,,
\bar V^{(3)}_c\geq \frac{i}{4}\,\Lambda\,,\\
\end{align*}
where only case I represents full inseparability. Figure \ref{fg:ent} (b) shows the minimum eigenvalue of $\bar V^{(3)}_j-\frac{i}{4}\,\Lambda$ for $j=a,b,c$. As the state propagates through the waveguide array, all the permuted correlation matrix violate the three-mode uncertainty relation and the test matrices have a negative eigenvalue. This means that we observe full inseparability through most of the propagation (case I). At normalized distances $\kappa z =(2n+1)\pi/2\sqrt{2}$ (where coefficients $T_{ab}$ and $T_{bc}$ are exactly zero) we have scenario II, meaning that waveguide $b$ is separable from the bipartite entangled reduced state of waveguides $a$ and $c$. At points $\kappa z = n\pi/\sqrt{2}$ (where $T_{ab}=T_{bc}=T_{ac}=0$) we recover the separable state of the input, corresponding to the case IV.

\section{Selecting squeezing multimodality by light polarization}
\label{S_polarization}

Elliptical waveguides allow to tune the coupling constant $\kappa$ by varying the input light polarization. These are made by a fs-laser-writing fabrication procedure, where the shape of the writing beam and the writing speed are adjusted to create waveguides with an elliptical cross-section~\cite{fs}. Highly elliptical waveguides \cite{pols}, whose cross-sections are $4\times 12$ $\mu$m$^2$ and separated by $23$ $\mu$m, can have a ratio between coupling constants for $H$- and $V$- polarized light close to 2. The considerable difference between coupling constants suggests that the evolution of an state can be radically different depending upon its polarization.

\begin{figure}[!bt]
 \begin{center}
  \includegraphics[width=.48\textwidth]{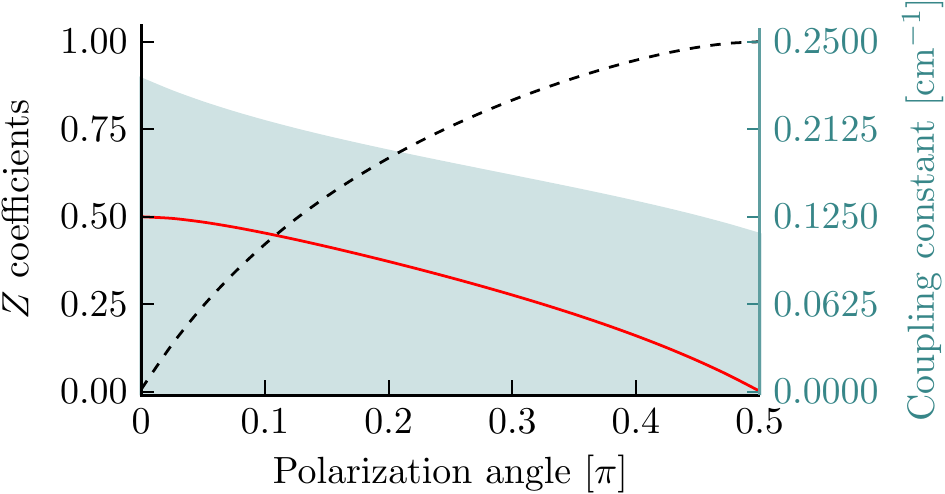}
 \end{center}
 \caption{$Z_a=Z_b$ (solid red curve) and $Z_{ab}$ (dashed black curve) as a function of the input light polarization angle in an optical dimer with elliptical waveguides. The shaded region and the right vertical axis show the range of achievable coupling constants as a function of the polarization angle~\cite{pols}.}\label{fg:polarization}
\end{figure}

We compute the evolution of the field through an optical dimer with elliptical waveguides, with tunable coupling constants as a function of the input light polarization angle. Following the mathematical treatment in Sec. \ref{S_TMG}, we assume a two single-mode squeezed states of equal squeezing parameter $\xi$ at the input. Figure \ref{fg:polarization} shows realistic values for the coupling constant as a function of the polarization angle. We compute the $Z$-coefficients for single-mode and two-mode squeezing at the output of the waveguides for a propagation distance $\kappa_Hz_\text{out}=\pi/2$. Notice how the multimodality of the squeezing varies drastically with the polarization angle. When light is horizontally polarized (0\textordmasculine) the output corresponds to only a single-mode squeezing. At 90\textordmasculine, $|Z_a|=|Z_b|=0$ and $|Z_{ab}|$ reaches its maximum value. Hence, it is possible to modify the output state from single-mode squeezing to sole two-mode squeezing, just by tuning the polarization angle of the input linear.  This suggests that a dimer built with elliptical waveguides can be used as a powerful tool for controlling and selecting the multi-modality of the squeezed light field. 

Squeezing tunability as a function of the light polarization may also offer the possibility of creating entanglement between the output light polarization and the order of the multimode squeezing. In this way, the order of the multimode squeezing can be chosen at the output by post-selecting in polarization. The calculation of this effect needs a more complicated mathematical description that will be evaluated in a separate work.

\section{Squeezing degradation in a lossy waveguide array}\label{S_losses} 

Optical losses in dielectric media are unavoidable, reducing the degree of squeezing of the propagating light. In general, any type of losses can be modeled with an effective beam splitter, where a fraction of the field is lost at one of the output ports. In such model the transmitted field is scaled by a factor $\eta$. The reduction in the squeezing as a function of the losses is given by~\cite{Mandel,Alberto}
\begin{equation}
S_{out}=-\frac{1}{2}\text{ln}\left[\eta(z)e^{-2S_{in}}+(1-\eta(z))\right],
\label{eq:loss}
\end{equation}
where $S_{in(out)}$ is the degree of squeezing at the input (output) and $\eta(z)$ is the loss after propagating a distance $z$. The variable $\eta$ can represent any type of losses in the system, such as losses by coupling the light into the waveguide or losses during propagation.

\begin{figure}[!bt]
\includegraphics[width=.47\textwidth]{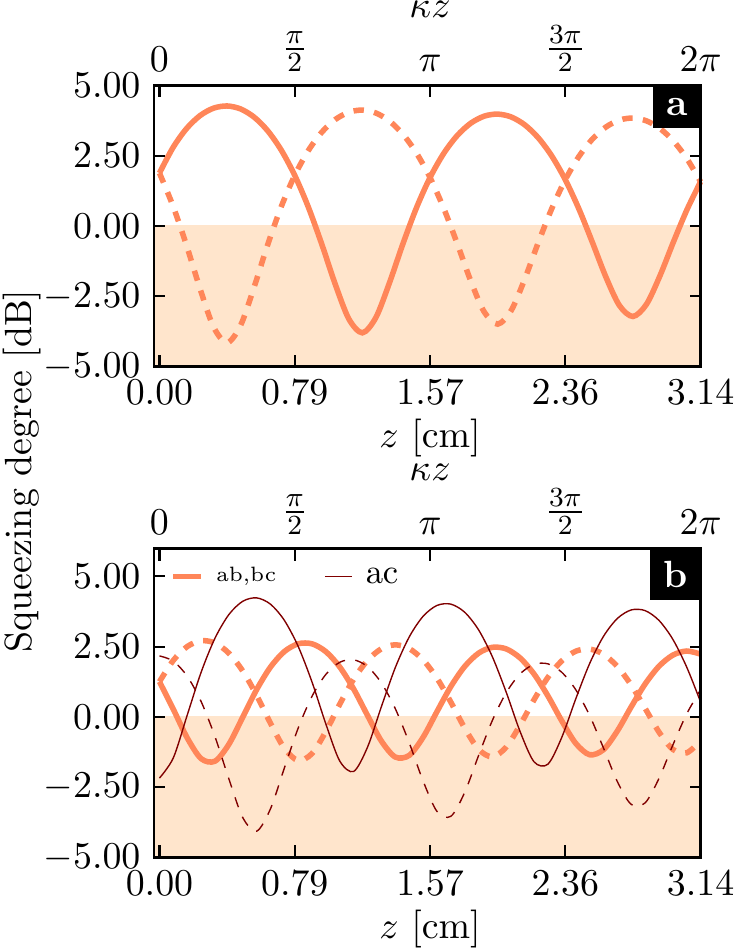}
 \caption{Degree of squeezing versus propagation distance. (a) Optical dimer for an input of two single-mode squeezed states with squeezing parameter $\xi_a=\xi_b=0.5$. The degree of squeezing is shown for the two-mode quadratures $\langle(\Delta X_1^{\text{2M}})^2\rangle$ and $\langle(\Delta X_2^{\text{2M}})^2\rangle$ (solid and dashed orange curves). (b) Optical trimer for an input state of three vacuum-squeezed states with squeezing parameter $\xi_a=\xi_c=0.25$ and $\xi_b=0.5$. The degree of squeezing is shown for the two-mode quadratures $X_1^{ab}=X_1^{bc}$ and $X_2^{ab}=X_2^{bc}$ (solid and dashed orange curves), as well as $X_1^{ac}$ and $X_2^{ac}$ (solid and dashed brown curves). For both figures the losses are $0.3$ dB$/$cm \cite{perdidas} and the evanescent coupling coefficient is $\kappa=2\,$cm$^{-1}$ \cite{fs}, which can be understood as an evolution of $\kappa z$ from $0$ to $2\pi$. Squeezing is observed when the curves are within the orange region.}  \label{fg:perdidas}
\end{figure}

Light squeezing, as well as its intensity, is attenuated exponentially as a function of propagation distance, following Beer-Lambert-Bouguer law. Assuming that losses affect equally all waveguides, the evolution of the state remains unchanged except for an overall reduction of the correlations as light attenuates. Figure \ref{fg:perdidas} shows the degradation of the degree of squeezing as a function of the propagation distance for an optical dimer and trimer. In both cases, squeezing is preserved despite of degradation. For realistic parameters it is possible to observe several oscillations of the degree of squeezing before it is completely attenuated. 

The degradation of squeezing can also be calculated by solving the master equation including losses, where photon absorption is described as an amplitude damping channel \cite{Santiago,Nielsen}. With this approach we find results equivalent to Fig. \ref{fg:perdidas}. However, Eq. (\ref{eq:loss}) also allows for easily including the effect of injection losses when coupling free-space propagation light into the waveguide. Injection losses are not considered in Fig. \ref{fg:perdidas}, but they will contribute to an overall reduction of the degree of squeezing. Nonetheless, we emphasize that multimode squeezing in a coupled waveguide array is preserved under common mechanism of losses.

\section{Conclusions and perspectives}
\label{S_conclusions}

We propose a method for the generation and manipulation of two- and three-mode squeezed states via the injection of single mode squeezed light in a linear array of two and three evanescently coupled waveguides. We observe that the squeezing evolves as it propagates through the waveguides exchanging the roles of single-mode and multimode squeezing. During light propagation, the waveguide array generates genuine multipartite entanglement. This offers a way to control the quantum non-linear property of multimode squeezing only using a linear optics element. We show that the order of multimode squeezing at the output of the system can be selected by choosing the polarization angle of the input light, offering a novel tuning knob for controlling squeezing. The parameters' tunability in an array of waveguides with linear response facilitates the engineering of quantum states of light. A review of realistic parameters for optical losses in a system of coupled waveguides shows that a significant degree of squeezing is preserved during propagation.

Waveguide arrays are a suitable platform to manipulate quantum light properties, such as squeezing and entanglement. They open new possibilities to generate and control multimode squeezing, with the potential to generate $N$-modes squeezed states with $N$ coupled waveguides. Multiple modes can be used as multiple probes that measure different local conditions, presenting a problem of multi-variables estimation. It is known that quantum correlations can improve the sensitivity of such scenario, making multimode squeezing a potential resource to improve multiple-variable measurements. By combining quantum light with photonic crystals, our results bring different insights and tools in the fields of quantum information and precision measurement.

\section*{Acknowledgments}
The authors acknowledge A. Marino, T. Woodworth, R. Vicencio, and D. Guzman for useful comments and
discussions. This work was supported in part by CONICYT Grant No. PAI 77180003, U-Inicia VID Grant No.
UI 004/2018, FONDECYT Grants No. 3180153 and No. 3180752, and Programa ICM Millennium Institute for
Research in Optics (MIRO). E.B. and C.M. contributed equally to this work.

\appendix

\section{Variance of the field in an optical dimer}\label{ap:quadrature}

We notice in Fig.\ref{fg:dima}(b) that for the single-mode quadrature variances, the squeezing is lost before $Z_a=Z_b=0$, because $Z_{ab}\neq0$. We can evaluate this using the following transformations ($\xi_i=r_i\exp{i\mu_i}$):

\begin{subequations}
 \begin{align}
 \label{TUGausson}
 S_a^{-1}S_b^{-1}UaU^{-1}S_{b}S_{a}&=[a\cosh({r_a})-a^\dagger e^{i\mu_a}\sinh({r_a})]\cos(\kappa z) \nonumber\\
 &-i[b\cosh({r_b})-b^\dagger e^{i\mu_b}\sinh({r_b})]\sin(\kappa z)\,\\
 S_a^{-1}S_b^{-1}UbU^{-1}S_{b}S_{a}=&i[a^\dagger e^{i\mu_a}\sinh({r_a})-a\cosh({r_a})]\sin(\kappa z)\nonumber\\
 &-[b\cosh({r_b})-b^\dagger e^{i\mu_b}\sinh({r_b})]\cos(\kappa z)\,.
 \end{align}
\end{subequations}

With this, the single-mode quadrature variances $\langle(\Delta X_j)^2\rangle$ can be written as

\begin{subequations}
\begin{align}
 \langle(\Delta X_1^a)^2\rangle = \frac{1}{4}&\left[-\cos^2(\kappa z)\sinh (2r_a)\,\cos(\mu_a)\right.\nonumber\\
 &+\sin^2(\kappa z)\sinh(2r_b)\,\cos(\mu_b)\nonumber\\
 &\left.+\cos^2(\kappa z)\cosh (2r_a)+\sin^2(\kappa z)\cosh^2(2r_b)  \right]\,,\\
  \langle(\Delta X_2^a)^2\rangle = \frac{1}{4}&\left[\cos^2(\kappa z)\sinh (2r_a)\,\cos(\mu_a)\right.\nonumber\\
 &-\sin^2(\kappa z)\sinh (2r_b)\,\cos(\mu_b)\nonumber\\
 &\left.+\cos^2(\kappa z)\cosh (2r_a)+\sin^2(\kappa z)\cosh^2 (2r_b)  \right]\,.
 \end{align}
 \label{Eq:A2}
\end{subequations}

These variances are the same for the second waveguide (mode ${b}$). For the particular case of $\xi_a=\xi_b$ real ($\mu_a=\mu_b=0$), Eqs. (\ref{Eq:A2}) reduce to $(1/4)[\cosh^2({2r_a})-\sinh({2r_a})\cos({2\kappa z})]$, plotted in the cyan curves in Fig.\ref{fg:dima}(b).

\section{Injection of a two-mode squeezed state to the optical dimer}\label{ap:TMSS}

 In the case of injecting a sole two-mode squeezed state to the system with two input and two output ports, the input state can be written as $\ket{\psi_0'}=S_{ab}\ket{0,0}$, with $S_{ab}=\exp\left \{(\chi a^\dagger b^\dagger-\chi^*ab)\right \}$ and $\chi=r_{ab}\exp{i\mu_{ab}}$. Then, the evolution of the state $\ket{\psi}=U^{-1}S_{ab}U\ket{0,0}$, has the form
\begin{equation}
\label{2mInjectionSolution}
 \ket{\psi}=\exp\left \{ \phi_a^*a+\phi_{b}^*b+ \phi_{ab}a^\dagger b^\dagger -H.c.\right \} \ket{0,0}
\end{equation}
with
\begin{subequations}\label{eq:2msolution}
 \begin{align}
  \phi_a& = \chi e^{i\delta} \cos(\theta)\sin(\theta)\,,\\
  \phi_b& =  -\chi e^{-i\delta} \cos(\theta)\sin(\theta)\,,\\
  \phi_{ab}&= \chi \left[\cos^2(\theta)-\sin^2(\theta)\right]\,.
 \end{align}
\end{subequations}

It is clear from the above expressions that in order to generated single-mode squeezed states $\phi_{ab}=0$, which implies that $\theta=\pi/4$. Then $\phi_a=\chi \exp{i\delta}/2$ and $\phi_b=-\chi\exp{-i\delta}/2$. A sole two-mode squeezed state will be recovered for $\theta=(2n+1)\pi/2$ or  $\theta=n\pi$. This general results is valid for the optical dimer with $\delta=\pi/2$ and $\theta=\kappa z$. 

\section{Three-mode squeezing in an optical trimer: general case}\label{apendice}

As we see in Sec. \ref{S_trimer}, an optical system with three input and three output ports can generate a state with mixed characteristics between three uncoupled single-mode squeezed states and a sole three-mode squeezed state
\begin{equation}
\label{apn:3mgausson}
\begin{split}
 \ket{\Psi}_T=&\exp\left \{\frac{1}{2}(T^*_aa^2+T^*_bb^2+T^*_cc^2\right.\\
 &\left.+T_{ab}a^\dagger b^\dagger+T_{ac}a^\dagger c^\dagger+T_{bc}b^\dagger c^\dagger- H.c.)\right \} \ket{0,0,0}.
 \end{split}
\end{equation}
This state can be generated by
\begin{equation}
\label{apn:3mgausson2}
 \ket{\Psi}_T=U_T^{-1}S_aS_bS_cU_T\ket{0,0,0},
\end{equation}
where $S_j$ are single-mode squeezing operators for $j$ different modes while $U$ represents a unitary rotation operator that coherently mixes the three modes $a$, $b$, and $c$,
\begin{equation}
\label{apn:3rotation}
U_T=\exp\left \{(\alpha^*ab^\dagger-\alpha ba^\dagger+\beta^*bc^\dagger-\beta cb^\dagger)\right \},
\end{equation}
with $\alpha=\theta_{\alpha}e^{i\delta_{\alpha}}$ and $\beta=\theta_{\beta}e^{i\delta_{\beta}}$ complex numbers. This is the most general expression for such a rotation. In order to obtain an analytical expression of  (\ref{apn:3mgausson}), we need to calculate $U_T^{-1}aU_T$, $U_T^{-1}bU_T$ and $U_T^{-1}cU_T$. We use the following recursive notation $[  A,   B]_n = [  A, [  A,   B]_{n-1}],$ where for $n=1$, and we define $[  A,   B]_1 = [  A,   B]$, to write the Baker-Campbell-Hausdorff formula as
\begin{equation}
     e^{  A}   B e^{-  A} =   B + \sum_{n=1}^{\infty}\frac{[  A,   B]_n}{n!}.
     \label{BCH}
\end{equation}

Given Eqs. (\ref{apn:3mgausson2}) and (\ref{apn:3rotation}), the operator $  A$ is
\begin{align}
\label{A_trans}
      A = \alpha^{\ast}  a  b^{\dagger}-\alpha   a^{\dagger}  b+\beta^{\ast}  b   c^{\dagger} -\beta   b^{\dagger}  c,
\end{align}
and the commutative relations that we need to calculate are
\begin{align}
    [  A,   a] &= \alpha   b,\\
    [  A,   b] &= \beta   c-\alpha^{\ast}   a,\\
    [  A,   c] &= -\beta^{\ast}  b.
\end{align}
Defining $\lambda^ 2 = |\alpha |^ 2 + | \beta |^ 2 $ we get for $\alpha\neq\beta$
\begin{align*}
      U_T^{-1}  a   U_T &=  \left( 1-2\frac{\theta_{\alpha}^2}{\lambda^2}\sin^2{\left(\frac{\lambda}{2}\right)}\right)   a +\frac{\theta_{\alpha}}{\lambda}e^{i\delta_{\alpha}}\sin{(\lambda)}   b\\
    &+2\frac{\theta_{\alpha}\theta_{\beta}}{\lambda^2}e^{i(\delta_{\alpha}+\delta_{\beta})}\sin^2{\left( \frac{\lambda}{2}\right)}  c,\\
    U_T^{-1}  b   U_T & = -\frac{\theta_{\alpha}}{\lambda}e^{-i\delta_{\alpha}}\sin{(\lambda)}  a + \cos{(\lambda)}  b +\frac{\theta_{\beta}}{\lambda}e^{i\delta_{\beta}}\sin{(\lambda)}  c,\\
       U_T^{-1}  c   U_T &=2\frac{\theta_{\alpha}\theta_{\beta}}{\lambda^2}e^{-i(\delta_{\alpha}+\delta_{\beta})} \sin^2{\left( \frac{\lambda}{2}\right)}  a -\frac{\theta_{\beta}}{\lambda}e^{-i\delta_{\beta}}\sin{(\lambda)}  b\\
     &+\left(1- 2\frac{\theta_{\beta}^2}{\lambda^2}\sin^2{\left( \frac{\lambda}{2}\right)} \right)   c,
\end{align*}

In the particular case of $|\alpha|=|\beta|=\theta$, and $\delta_\alpha=\delta_\beta=\delta$, i.e. the same coupling between waveguides $a$ and $b$ and $b$ and $c$, we can reduce the above expressions to
\begin{align*}
  U_T^{-1}  a   U_T &= \cos^2{\left( \frac{\theta}{\sqrt{2}}\right)}   a +\frac{e^{i\delta}}{\sqrt{2}}\sin{\left(\sqrt{2}\theta\right)}   b+e^{2i\delta}\sin^2{\left(\frac{\theta}{\sqrt{2}}\right)}  c,\\
  U_T^{-1}  b   U_T & =-\frac{1}{\sqrt{2}}e^{-i\delta}\sin{\left(\sqrt{2}\theta\right)}  a + \cos{\left(\sqrt{2}\theta \right)}  b\\
 & +\frac{1}{\sqrt{2}}e^{i\delta}\sin{\left(\sqrt{2}\theta\right)}  c\\
      U_T^{-1}  c   U_T &=e^{-2i\delta} \sin^2{\left( \frac{\theta}{\sqrt{2}}\right)}  a\\
    & -\frac{1}{\sqrt{2}}e^{-i\delta}\sin{\left(\sqrt{2}\theta\right)}  b+\cos^2{\left( \frac{\theta}{\sqrt{2}}\right)}   c.
\end{align*}

Using these results and calculating $U_T^{\dagger}u^2U_T$ and $U_T^{\dagger} u^{\dagger 2}U_T$, with $u=a,b,c$, we find the following $T_i$ and $T_{ij}$ coefficients for $\alpha\neq\beta$ 

\begin{widetext}
\begin{equation}
\label{eqA1}
\begin{split}
 T_a & = \xi_a\left( 1-2\frac{\theta_{\alpha}^2}{\lambda^2}\sin^2{\left(\frac{\lambda}{2}\right)}\right)^2 +\xi_b\frac{\theta_{\alpha}^2}{\lambda^2}e^{2i\delta_{\alpha}}\sin^2{(\lambda)}+\xi_c4\frac{\theta_{\alpha}^2\theta_{\beta}^2}{\lambda^4}e^{2i(\delta_{\alpha}+\delta_{\beta})} \sin^4{\left( \frac{\lambda}{2}\right)}\\
 T_b& = \xi_a \frac{\theta_{\alpha}^2}{\lambda^2}e^{-2i\delta_{\alpha}}\sin^2{(\lambda)}+\xi_b\cos^2{(\lambda)}+\xi_c\frac{\theta_{\beta}^2}{\lambda^2}e^{2i\delta_{\beta}}\sin^2{(\lambda)} \\ 
 T_c& =\xi_a 4\frac{\theta_{\alpha}^2\theta_{\beta}^2}{\lambda^4}e^{-2i(\delta_{\alpha}+\delta_{\beta})}\sin^4{\left( \frac{\lambda}{2}\right)}+\xi_b\frac{\theta_{\beta}^2}{\lambda^2}e^{-2i\delta_{\beta}}\sin^2{(\lambda)}+\xi_c\left(1- 2\frac{\theta_{\beta}^2}{\lambda^2}\sin^2{\left( \frac{\lambda}{2}\right)} \right)^2\\
 T_{ab} & = -\left\{ \xi_a 2\frac{\theta_{\alpha}}{\lambda}e^{-i\delta_{\alpha}}\left( 1-2\frac{\theta_{\alpha}^2}{\lambda^2}\sin^2{\left(\frac{\lambda}{2}\right)}\right)\sin{(\lambda)}-\xi_b \frac{\theta_{\alpha}}{\lambda}e^{i\delta_{\alpha}}\sin{(2\lambda)}-\xi_c 4\frac{\theta_{\alpha}\theta_{\beta}^2}{\lambda^3}e^{i(\delta_{\alpha}+2\delta_{\beta})} \sin{(\lambda)}\sin^2{\left( \frac{\lambda}{2}\right)} \right\}\\
  T_{ac} & = -\left\{\xi_a4\frac{\theta_{\alpha}\theta_{\beta}}{\lambda^2}e^{-i(\delta_{\alpha}+\delta_{\beta})}\left( 1-2\frac{\theta_{\alpha}^2}{\lambda^2}\sin^2{\left(\frac{\lambda}{2}\right)}\right)\sin^2{\left( \frac{\lambda}{2}\right)}-\xi_b 2\frac{\theta_{\alpha}\theta_{\beta}}{\lambda^2}e^{i(\delta_{\alpha}-\delta_{\beta})}\sin^2{(\lambda)}\right.\\
&\left.+\xi_c 4\frac{\theta_{\alpha}\theta_{\beta}}{\lambda^2}e^{i(\delta_{\alpha}+\delta_{\beta})} \left(1- 2\frac{\theta_{\beta}^2}{\lambda^2}\sin^2{\left( \frac{\lambda}{2}\right)} \right)\sin^2{\left( \frac{\lambda}{2}\right)}\right\}\\
T_{bc} & = -\left\{\xi_a 4\frac{\theta_{\alpha}^2\theta_{\beta}}{\lambda^3}e^{-i(2\delta_{\alpha}+\delta_{\beta})}\sin{(\lambda) }\sin^2{\left( \frac{\lambda}{2}\right)}+\xi_b\frac{\theta_{\beta}}{\lambda}e^{-i\delta_{\beta}}\sin{(2\lambda)}-\xi_c 2\frac{\theta_{\beta}}{\lambda}e^{i\delta_{\beta}}\left(1- 2\frac{\theta_{\beta}^2}{\lambda^2}\sin^2{\left( \frac{\lambda}{2}\right)} \right)\sin{(\lambda)}\right\}.
 \end{split}
\end{equation}

For equal coupling ($\alpha=\beta$), we obtain the following simplified expressions
\begin{equation}
\label{eqA2}
\begin{split}
T_a&= \xi_a\cos^4{\left( \frac{\theta}{\sqrt{2}}\right)}+\xi_b\frac{1}{2}e^{2i\delta}\sin^2{\left(\sqrt{2}\theta\right)}+\xi_ce^{4i\delta}\sin^4{\left(\frac{\theta}{\sqrt{2}}\right)} \\
T_b&= \xi_a\frac{1}{2}e^{-2i\delta}\sin^2{\left( \sqrt{2}\theta\right)}+\xi_b\cos^2{\left( \sqrt{2}\theta\right)}+\xi_c\frac{1}{2}e^{2i\delta}\sin^2{\left(\sqrt{2}\theta\right)} \\
T_c&= \xi_c e^{-4i\delta}\sin^4{\left( \frac{\theta}{\sqrt{2}}\right)}+\xi_b\frac{1}{2}e^{-2i\delta}\sin^2{\left( \sqrt{2}\theta\right)}+\xi_c\cos^4{\left( \frac{\theta}{\sqrt{2}}\right)} \\
T_{ab} &=-\left\{\xi_a \sqrt{2}e^{-i\delta}\sin{\left( \sqrt{2}\theta\right)}\cos^2{\left( \frac{\theta}{\sqrt{2}}\right)}-\xi_b\frac{1}{\sqrt{2}}e^{i\delta}\sin{\left(2\sqrt{2}\theta \right)}-\xi_c\sqrt{2}e^{3i\delta}\sin{\left( \sqrt{2}\theta \right)}\sin^2{\left( \frac{\theta}{\sqrt{2}}\right)}\right\} \\
T_{ac} &=-\left\{\xi_a \frac{1}{2}e^{-2i\delta}\sin^2{\left( \sqrt{2}\theta\right) }-\xi_b\sin^2{\left( \sqrt{2}\theta\right)}+\xi_c\frac{1}{2}e^{2i\delta}\sin^2{\left( \sqrt{2}\theta\right)}\right\} \\
T_{bc} &= -\left\{\xi_a \sqrt{2}e^{-i3\delta}\sin{\left( \sqrt{2}\theta\right)}\sin^2{\left( \frac{\theta}{\sqrt{2}}\right)}+\xi_b \frac{1}{\sqrt{2}}e^{-i\delta}\sin{\left( 2\sqrt{2}\theta\right)}-\xi_c\sqrt{2}e^{i\delta}\sin{\left( \sqrt{2}\theta\right)}\cos^2{\left( \frac{\theta}{\sqrt{2}}\right)}\right\}.
\end{split}
\end{equation}

These are the coefficients that lead to Eq.(\ref{coeftrimero}) in the main text when making $\delta=\pi/2$.

\end{widetext}

\bibliography{References}
\bibliographystyle{unsrtnat}

\end{document}